\documentclass[11pt]{article}
\usepackage{setspace}
\usepackage{amssymb}

\usepackage{amsmath}
\usepackage{graphicx}
\usepackage{theorem}
\usepackage{here}
\usepackage{booktabs}

\usepackage{natbib}

\usepackage{color}
\usepackage{soul}
\usepackage{ulem}

\newtheorem{thm}{Theorem}
\newtheorem{lem}{Lemma}

\theorembodyfont{\rmfamily}

\def\al{{\alpha}}
\def\be{{\beta}}
\def\ga{{\gamma}}
\def\de{{\delta}}
\def\ep{{\varepsilon}}
\def\la{{\lambda}}
\def\si{{\sigma}}
\def\om{{\omega}}

\def\ta{{\tau}}

\def\bbe{{\text{\boldmath $\beta$}}}

\def\bta{{\text{\boldmath $\eta$}}}

\def\bep{{\text{\boldmath $\varepsilon$}}}

\def\bth{{\text{\boldmath $\theta$}}}

\def\bpsi{{\text{\boldmath $\psi$}}}

\def\sih{{\hat \si}}
\def\psih{{\widehat \psi}}
\def\tah{{\hat \tau}}

\def\lah{{\hat \la}}

\def\delt{{\tilde \delta}}

\def\lat{{\tilde \la}}

\def\sit{{\tilde \si}}

\def\bbeh{{\widehat \bbe}}

\def\btah{{\widehat \bta}}

\def\bthh{{\widehat \bth}}

\def\bpsih{{\widehat \bpsi}}

\def\bbet{{\widetilde \bbe}}
\def\btat{{\widetilde \bta}}

\def\De{{\Delta}}
\def\Si{{\Sigma}}

\def\La{{\Lambda}}

\def\bSi{{\text{\boldmath $\Si$}}}

\def\bLa{{\text{\boldmath $\La$}}}

\def\bSit{{\widetilde \bSi}}

\def\a{{\text{\boldmath $a$}}}
\def\b{{\text{\boldmath $b$}}}

\def\u{{\text{\boldmath $u$}}}

\def\x{{\text{\boldmath $x$}}}
\def\y{{\text{\boldmath $y$}}}

\def\A{{\text{\boldmath $A$}}}
\def\B{{\text{\boldmath $B$}}}
\def\C{{\text{\boldmath $C$}}}

\def\G{{\text{\boldmath $G$}}}

\def\I{{\text{\boldmath $I$}}}
\def\J{{\text{\boldmath $J$}}}

\def\P{{\text{\boldmath $P$}}}

\def\R{{\text{\boldmath $R$}}}

\def\X{{\text{\boldmath $X$}}}

\def\Z{{\text{\boldmath $Z$}}}

\def\Po{{\overline P}}

\def\Yo{{\overline Y}}

\def\bh{{\hat b}}

\def\bbh{{\hat \b}}
\def\ybt{{\tilde \y}}

\def\bbt{{\tilde \b}}
\def\bept{{\tilde \bep}}

\def\Xbt{{\widetilde \X}}

\def\Zbt{{\widetilde \Z}}

\def\Pbt{{\widetilde \P}}

\def\Rbt{{\widetilde \R}}
\def\bSit{{\widetilde \bSi}}
\def\Rt{{\widetilde R}}

\def\Jc{{\cal J}}

\def\Nc{{\cal N}}

\def\Pc{{\cal P}}

\def\tr{{\rm tr\,}}
\def\diag{{\rm diag\,}}
\def\rank{{\rm rank\,}}

\def\zero{{\bf\text{\boldmath $0$}}}
\def\one{{\bf\text{\boldmath $1$}}}
\def\dd{{\rm d}}

\def\non{{\nonumber}}

\def\cAI{{\rm cAI}}

\def\IC{{\rm IC}}
\def\MSE{{\rm MSE}}

\date{}

\begin{document}
\title{Conditional Akaike information under covariate shift with application to small area estimation\footnote{Last update: \today}}

\author{
Yuki Kawakubo\thanks{Graduate School of Social Sciences, Chiba University (E-mail: kawakubo@chiba-u.jp)}
 \ \ Shonosuke Sugasawa\thanks{Risk Analysis Research Center, The Institute of Statistical Mathematics} \
 and Tatsuya Kubokawa\thanks{Faculty of Economics, University of Tokyo}
}
\maketitle
\begin{abstract}

In this study, we consider the problem of selecting explanatory variables of fixed effects in linear mixed models under covariate shift, which is when the values of covariates in the model for prediction differ from those in the model for observed data.
We construct a variable selection criterion based on the conditional Akaike information introduced by \citet{VB05}.
We focus especially on covariate shift in small area estimation and demonstrate the usefulness of the proposed criterion.
In addition, numerical performance is investigated through simulations, one of which is a design-based simulation using a real dataset of land prices.

\par\vspace{4mm}
\end{abstract}

\section{Introduction}
\label{sec:int}

Linear mixed models have been rigorously studied and have many applications, for example, longitudinal data analysis in biostatistics, panel data analysis in econometrics, and small area estimation in official statistics.
The problem of selecting explanatory variables in linear mixed models is important and has been investigated in the literature.
\citet{MSW13} provide a good survey of model selection for linear mixed models, which includes information criteria, shrinkage methods such as LASSO, the fence method \citep{JRGN08} and Bayesian methods.

When the purpose of variable selection is to identify a set of significant variables that provide good prediction, Akaike information criteria \citep{Aka73,Aka74} are often used.
However, the Akaike information criterion (AIC) is based on the marginal likelihood, the likelihood with the random effects integrated out, and is not appropriate when the prediction is focused on random effects.
Then, \citet{VB05} proposed considering the Akaike information based on the conditional density given the random effects and proposed the conditional AIC (cAIC).
Here, we provide a brief explanation of the cAIC concept: let $f(\y|\b,\bta, \X, \Z)$ be the conditional density function of $\y$ given $\b$, where $\y$ is an observable random vector of the response variable, $\b$ is a random vector of random effects, $\bta = (\bbe^\top,\si^2)^\top$ is a vector of unknown fixed parameters $\bbe$ and an unknown variance component $\si^2$, and $\X$ and $\Z$ are design matrices of covariates and random effects, respectively.
In our setup, although $\X$ and $\Z$ are neither random nor unknown parameters, we emphasize that the distribution of $\y$ is characterized by $\X$ and $\Z$.
This is for ease of notation when we consider the situation of covariate shift later.
Let $\pi(\b|\si^2)$ be the density function of $\b$.
\citet{VB05} proposed measuring the prediction risk of the plug-in predictive density $f(\ybt|\bbh,\btah, \X, \Z)$ using Kullback--Leibler divergence:
\begin{align}
\iint \left[ \int \log \left\{ \frac{f(\ybt|\b, \bta, \X, \Z)}{f(\ybt|\bbh, \btah, \X, \Z)} \right\} f(\ybt|\b,\bta, \X, \Z) \dd\ybt \right] f(\y|\b,\bta, \X, \Z) \pi(\b|\si^2) \dd\y\dd\b, \label{eqn:cKL}
\end{align}
where $\ybt$ is an independent replication of $\y$ given $\b$, and $\bbh$ and $\btah$ are predictors or estimators of $\b$ and $\bta$, respectively.
The cAIC is an (asymptotically) unbiased estimator of a part of the risk in (\ref{eqn:cKL}), which is called the conditional Akaike information (cAI), given by
$$
{\rm cAI}=-2\iiint \log \left\{ f(\ybt|\bbh,\btah,\X, \Z) \right\} f(\ybt|\b,\bta,\X, \Z) f(\y|\b,\bta,\X, \Z) \pi(\b|\si^2) \dd\ybt\dd\y\dd\b.
$$
cAIC as a variable selection criterion in linear mixed models has been studied by \citet{LWZ08}, \citet{GK10}, \citet{SK10}, \citet{Kub11}, \citet{KK14}, and others.
Furthermore, the cAIC has been constructed as a variable selection criterion in generalized linear mixed models by \citet{DOXV11}, \citet{YY12}, \citet{YZY13}, \citet{SKWG14}, and others.

Consider the prediction problem where it is often the case that the covariate values in the model for prediction are different from those in the model for the observed data.
We call this covariate shift and refer the model in which $\y$ is the vector of response variables as the ``model for the observed data'', and the model in which $\ybt$ is the vector of response variables as the ``model for prediction''.
Note that the term ``covariate shift'' was first used by \citet{Shi00}, who defined it as the situation in which the distribution of the covariates in the model for prediction differs from that in the model for observed data.
In this study, although we treat the covariates as non-random, we still use the term ``covariate shift''.
Even when information about covariates in the model for prediction can be used, most Akaike-type information criteria do not use it.
This is because it is assumed that the model for prediction is the same as the model for the observed data when deriving such criteria.
For cAIC, the conditional density of $\y$ given $\b$ and that of $\ybt$ given $\b$ are the same, both of which are denoted by $f(\cdot|\b, \bta, \X, \Z)$.
On the other hand, under the covariate shift, the conditional density of $\ybt$ given $\b$ is different from that of $\y$ given $\b$ and is denoted by $g(\ybt|\b, \bta, \Xbt, \Zbt)$.
Note that the distribution of $\ybt$ given $\b$ is characterized by $\Xbt$ and $\Zbt$, not $\X$ and $\Z$.
When the aim of variable selection is to choose the best model for prediction, it is not appropriate to use the covariates that are only in the model for the observed data.
Therefore, we redefine the cAI under covariate shift, as follows:
$$
{\rm cAI}=-2\iiint \log \left\{ g(\ybt|\bbh,\btah,\Xbt, \Zbt) \right\} g(\ybt|\b,\bta,\Xbt, \Zbt) f(\y|\b,\bta,\X, \Z) \pi(\b|\si^2) \dd\ybt\dd\y\dd\b,
$$
and construct an information criterion as an asymptotically unbiased estimator of cAI.
\citet{Sat97, Sat00} considered a similar problem with the multivariate linear regression model and proposed a variable selection criterion.
It is important to note that we do not assume that the candidate model is overspecified, in other words, that the candidate model includes the true model.
Although most Akaike-type information criteria make this assumption, it is not appropriate when estimating the cAI under covariate shift.
We discuss this further in Section \ref{subsec:drawback}.

As an important application of covariate shift, we focus on small area estimation, which is based on a finite super-population model.
We consider the situation in which we are interested in the finite subpopulation (area) mean of a particular characteristic and some of its values in the subpopulation are observed through a sampling procedure.
When the sample size in each area is small, the problem is referred to as small area estimation.
For details regarding small area estimation, see \citet{RM15}, \citet{DG12}, \citet{Pfe13}, and others.
The model-based approach in small area estimation often assumes that the finite population has a super-population with random effects and borrows strength from other areas to estimate (predict) the small area (finite subpopulation) mean.
The well-known unit-level model is the nested error regression model (NERM), which is a kind of linear mixed model, and was introduced by \citet{BHF88}.
The NERM can be used when the values of the auxiliary variables for the units with characteristics of interest (response variables in the model) are observed through survey sampling.
%\sout{In this context, the necessity of considering covariate shift means that not only the auxiliary variables for sampled units of characteristics of interest, but also those for non-sampled units or aggregated values of those for all units, are available.}
In this context, the ``model for the observed data'' is the NERM for sampled units.
On the other hand, two types of ``model for prediction'' can be considered.
One is the unit-level model, which can be used when the values of the auxiliary variables are known for all units.
The other is the area-level model, which can be when the mean of each auxiliary variable is known for each small area.
The latter is frequently the case in official statistics and the model introduced by \citet{FH79} is often used.

The rest of this paper is organized as follows.
In Section \ref{sec:setup}, we explain the setup of the variable selection problem.
In Section \ref{sec:criteria}, we define the cAI under covariate shift in linear mixed models and obtain an asymptotically unbiased estimator of the cAI.
In Section \ref{sec:ex}, we provide an example of covariate shift focused on small area estimation.
In Section \ref{sec:simu}, we investigate the numerical performance of the proposed criteria via a design-based simulation based on a real dataset of land prices.
We conclude this paper with some discussion in Section \ref{sec:dis}.
All proofs are given in the Appendix.

\section{The variable selection problem}
\label{sec:setup}

\subsection{The class of candidate models}

We focus on variable selection for fixed effects.
First, we consider the collection of candidate models as follows:
let the $n\times p_\om$ matrix $\X(\om)$ consist of all the explanatory variables and assume $\rank(\X(\om)) = p_\om$.
To define candidate models by the index set $j$, suppose that $j$ denotes a subset of $\om = \{ 1,\dots,p_\om \}$ containing $p_j$ elements, {\it i.e.}, $p_j = \#(j)$, and $\X(j)$ consists of $p_j$ columns of $\X(\om)$ indexed by the elements of $j$.
We define the class of candidate models by $\Jc = \Pc(\om)$, namely, the power set of $\om$, in which we call $\om$ the full model.
We assume that the true model exists in the class of candidate models $\Jc$, and denote it by $j_*$.
The dimension of the true model is $p_{j_*}$, which is abbreviated $p_*$.

We next introduce the terms ``overspecified'' and ``underspecified''.
Candidate model $j$ is overspecified if $j_* \subseteq j$ following \citet{FS97} or \citet{KK14}.
The set of overspecified models is denoted by $\Jc_+ = \{ j\in \Jc | j_* \subseteq j \}$.
On the other hand, candidate model $j$ is underspecified when $j_* \not\subseteq j$.
The set of underspecified models is denoted by $\Jc_- = \Jc \setminus \Jc_+$.
It is important to note that most Akaike-type information criteria are derived under the assumption that the candidate model is overspecified.
However, the assumption is not appropriate when considering the covariate shift, which is explained in Section \ref{subsec:drawback}.
Thus, we derive the criterion without the overspecified assumption.

In the following two subsections, we describe the model for the observed data and the model for prediction.

\subsection{Model for the observed data}

The candidate model $j$ for the observed data is the linear mixed model
\begin{equation}
\label{eqn:omodel}
\y=\X(j)\bbe_j +\Z\b_j +\bep_j,
\end{equation} 
where $\y$ is an $n\times 1$ observation vector of response variables, $\X(j)$ and $\Z$ are $n\times p_j$ and $n\times q$ matrices of covariates, respectively, $\bbe_j$ is a $p_j \times 1$ vector of regression coefficients, $\b_j$ is a $q\times 1$ vector of random effects, and $\bep_j$ is an $n\times 1$ vector of random errors.
Let $\b_j$ and $\bep_j$ be mutually independent, $\b_j\sim\Nc_q(\zero,\si_j^2\G)$ and $\bep_j\sim\Nc_n(\zero,\si_j^2\R)$ where $\G$ and $\R$ are $q\times q$ and $n\times n$ positive definite matrices and $\si_j^2$ is an unknown scalar.
%We assume that $\G$ and $\R$ are known and $\si_j^2$ is unknown.
A common variance parameter $\si_j^2$ for the random effects $\b_j$ and the error terms $\bep_j$ is also assumed by \citet{VB05}, who introduced the original cAIC.
We assume that $\G$ and $\R$ are known when deriving the criteria, though $\G$ and $\R$ include unknown variance components, the vector of which is denoted by $\bpsi$, in many applications.
In practice, $\G(\bpsi)$ and $\R(\bpsi)$ are replaced with their plug-in estimators $\G(\bpsih)$ and $\R(\bpsih)$ for some consistent estimator $\bpsih$.
The effect of this replacement is limited because $\bpsi$ is a nuisance parameter vector when we are selecting only explanatory variables.
This strategy is adopted by many variable selection criteria, including Mallows' $C_p$ \citep{Mal73}, and the residual information criterion (RIC) introduced by \citet{ST02}.

The true model $j_*$ for the observed data is
\begin{equation*}
%\label{eqn:CS_tomodel}
\y = \X(\om)\bbe_* + \Z\b_* + \bep_*,
\end{equation*}
where $\b_* \sim \Nc_q(\zero,\si^2_*\G)$, $\bep_* \sim \Nc_n(\zero,\si^2_*\R)$ and $\bbe_*$ is a $p_\om \times 1$ vector of regression coefficients, whose $p_\om - p_*$ components are exactly $0$ and the rest nonzero.
The equation above implies that the true model exists in the class of candidate models $\Jc$.
Note that $\X(\om)$ is the $n\times p_\om$ matrix of covariates for the full model $\om$.
The marginal distribution of $\y$ is
$$
\y \sim \Nc_n(\X(\om)\bbe_*, \si^2_*\bSi),
$$
where $\bSi = \Z\G\Z^\top + \R$.
For the true model, the conditional density function of $\y$ given $\b_*$ and the density function of $\b_*$ are denoted by $f(\y|\b_*,\bta_*, \X(\om), \Z)$ for $\bta_* = (\bbe_*^\top,\si_*^2)^\top$ and $\pi(\b_*|\si^2_*)$, respectively.

\subsection{Model for prediction}

The candidate model $j$ for prediction is the linear mixed model, which has the same regression coefficients $\bbe_j$ and random effects $\b_j$ as the candidate model for observed data $j$, but different covariates, namely,
\begin{equation}
\label{eqn:pmodel}
\ybt = \Xbt(j)\bbe_j + \Zbt\b_j + \bept_j,
\end{equation}
where $\ybt$ is an $m \times 1$ random vector of the target of prediction, $\Xbt(j)$ and $\Zbt$ are $m\times p_j$ and $m\times q$ matrices of covariates whose columns correspond to those of $\X(j)$ and $\Z$, respectively, and $\bept_j$ is an $m \times 1$ vector of random errors, which is independent of $\b_j$ and $\bep_j$ and is distributed as $\bept_j \sim \Nc_m(\zero, \si_j^2\Rbt)$, where $\Rbt$ is a known $m\times m$ positive definite matrix.
We assume that we know the values of $\Xbt(j)$ and $\Zbt$ in the model for prediction and that they are not necessarily the same as those of $\X(j)$ and $\Z$ in the model for the observed data.
The conditional density function of $\ybt$ given $\b_j$ for the model $j$ is denoted by $g_j(\ybt|\b_j,\bta_j,\Xbt(j),\Zbt)$ for $\bta_j = (\bbe_j^\top,\si_j^2)^\top$.

The true model $j_*$ for prediction is
\begin{equation*}
%\label{eqn:CS_tpmodel}
\ybt = \Xbt(\om)\bbe_* + \Zbt\b_* + \bept_*,
\end{equation*}
where $\Xbt(\om)$ is an $m \times p_\om$ matrix of covariates and $\bept_* \sim \Nc_m(\zero,\si_*^2\Rbt)$.
The marginal distribution of $\ybt$ is
$$
\ybt \sim \Nc_m( \Xbt(\om)\bbe_*, \si^2_*\bSit ),
$$
where $\bSit = \Zbt\G\Zbt^\top + \Rbt$.
For the true model, the conditional density function of $\ybt$ given $\b_*$ is denoted by $g(\ybt|\b_*,\bta_*,\Xbt(\om), \Zbt)$.

\section{Proposed criterion}
\label{sec:criteria}

\subsection{Conditional Akaike information under covariate shift}

The cAI introduced by \citet{VB05} measures the prediction risk of the plug-in predictive density $g_j(\ybt|\bbh_j,\btah_j,\Xbt(j),\Zbt)$, where $\btah_j = (\bbeh_j^\top,\sih_j^2)^\top$ is the maximum likelihood estimator given by
\begin{align*}
\bbeh_j =& \ (\X(j)^\top\bSi^{-1}\X(j))^{-1}\X(j)^\top\bSi^{-1}\y, \\
\sih_j^2 =& \ (\y - \X(j)\bbeh_j)^\top \bSi^{-1} (\y - \X(j)\bbeh_j) / n,
\end{align*}
and $\bbh_j$ is the empirical Bayes estimator of $\b_j$ for quadratic loss, which is given by
$$
\bbh_j = \G\Z^\top\bSi^{-1}(\y - \X(j)\bbeh_j).
$$
The cAI under covariate shift is
\begin{align*}
\cAI =& \ E^{(\y,\b_*)}E^{\ybt|\b_*}\left[ -2\log \{ g_j(\ybt|\bbh_j,\btah_j,\Xbt(j),\Zbt) \} \right] \\
=& \ E^{(\y,\b_*)}E^{\ybt|\b_*} \left[ m\log(2\pi\sih_j^2) + \log|\Rbt| + (\ybt - \Xbt(j)\bbeh_j -\Zbt\bbh_j)^\top \Rbt^{-1}(\ybt - \Xbt(j))\bbeh_j -\Zbt\bbh_j) / \sih_j^2 \right]
\end{align*}
where $E^{(\y,\b_*)}$ and $E^{\ybt|\b_*}$ are expectations with respect to the joint distribution of $(\y,\b_*) \sim f(\y|\b_*,\bta_*,\X(\om),\Z) \pi(\b_*|\si_*^2)$ and the conditional distribution of $\ybt$ given $\b_*$, namely $\ybt|\b_* \sim g(\ybt|\b_*,\bta_*,\Xbt(\om),\Zbt)$, respectively.
Taking expectations with respect to $\ybt|\b_* \sim g(\ybt|\b_*,\bta_*,\Xbt(\om),\Zbt)$ and $\b_*|\y \sim \Nc_q(\bbt_*,\si_*^2(\G - \G\Z^\top\bSi^{-1}\Z\G))$ for $\bbt_* = \G\Z^\top\bSi^{-1}(\y - \X(\om)\bbe_*)$, we obtain
\begin{equation}
\label{eqn:CScAI}
\cAI = E\left[ m\log(2\pi\sih_j^2) + \log|\Rbt| + \tr(\Rbt^{-1}\bLa)\cdot \si_*^2 / \sih_j^2 + \a^\top\Rbt^{-1}\a / \sih_j^2 \right]
\end{equation}
where
\begin{equation*}
%\label{eqn:laa}
\begin{split}
\bLa =& \ \bSit -\Zbt\G\Z^\top\bSi^{-1}\Z\G\Zbt^\top, \\
\a =& \ (\Xbt(j)\bbeh_j -\Xbt(\om)\bbe_*) -\Zbt\G\Z^\top\bSi^{-1}(\X(j)\bbeh_j - \X(\om)\bbe_*).
\end{split}
\end{equation*}

\subsection{Drawback of the overspecified model assumption}
\label{subsec:drawback}

Most Akaike-type information criteria are derived under the assumption that the candidate model includes the true model, namely, the overspecified assumption.
Although the assumption seems inappropriate, the resulting criterion based on this assumption works well in practice.
This is because the likelihood part of the criterion is a naive estimator of the risk function, namely, the cAI in the context of the cAIC.

Under the covariate shift situation, however, we cannot rely on the likelihood part to be a good estimator of the cAI.
That is, the drawback of the overspecified assumption is more serious in the situation of covariate shift than the usual one.
In Section \ref{subsec:simu_bias}, we show that an unbiased estimator of the cAI under the overspecified assumption, $\widehat{\cAI}_u$ in (\ref{eqn:CScAIC}), has large bias for estimating the cAI of underspecified models.

Thus, we evaluate and estimate the cAI directly for both overspecified and underspecified models in the following subsection.

\subsection{Evaluation and estimation of cAI}

We evaluate cAI in (\ref{eqn:CScAI}) both for the overspecified model and for the underspecified model.
We assume that the full model $\om$ is overspecified, that is, the collection of the overspecified models $\Jc_+$ is not an empty set.
In addition, we assume that the size of the response variable in the model for prediction $m$ is of order $O(n)$.

When the candidate model $j$ is overspecified, $n\sih_j^2 / \si_*^2$ follows a chi-squared distribution and we can evaluate the expectation in (\ref{eqn:CScAI}) exactly.
However, for the underspecified model, this is not true.
In this case, we approximate asymptotically the cAI as per the following theorem.

\begin{thm}
\label{thm:cAIm}
For the overspecified case, it follows that ${\rm cAI} = E[m\log(2\pi\sih_j^2)] + \log |\Rbt| + R^*$, where
$$
R^* = {n\ga \over n-p_j-2},
$$
for $\ga = \tr(\Rbt^{-1}\bLa) + \tr[\Rbt^{-1}\A(\X(j)^\top\bSi^{-1}\X(j))^{-1}\A^\top]$ and $\A = \Xbt(j) - \Zbt\G\Z^\top\bSi^{-1}\X(j)$.
For the underspecified case, $\cAI$ is approximated as
\begin{equation}
\label{eqn:cAIapp}
{\rm cAI} = E[m\log(2\pi\sih_j^2)] + \log|\Rbt| + R^* + R_1 + R_2 + R_3 + R_4 + O(n^{-1}),
\end{equation}
where
\begin{align*}
R_1 =& \ \ga(\la - 1), \\
R_2 =& \ \ga \cdot n^{-1}\{ -2\la^3 + (p_j+4)\la^2 - (p_j+2) \}, \\
R_3 =& \ \la \cdot \bbe_*^\top\B^\top\Rbt^{-1}\B\bbe_* / \si_*^2, \\
R_4 =& \ n^{-1} \{ -2\la^3 + (p_j+4)\la^2 \} \times \bbe_*^\top \B^\top \Rbt^{-1}\B\bbe_* / \si_*^2,
\end{align*}
for
\begin{align*}
\la =& \ 1 / (1+\de), \quad \de = \bbe_*^\top\X(\om)^\top(\P_\om - \P_j)\X(\om)\bbe_* / (n\si_*^2), \\
\B =& \ \Pbt_j\X(\om) - \Xbt(\om) + \Zbt\G\Z^\top(\P_\om - \P_j)\X(\om) , \\
\P_j =& \ \bSi^{-1}\X(j)(\X(j)^\top\bSi^{-1}\X(j))^{-1}\X(j)^\top\bSi^{-1}, \\
\P_\om =& \ \bSi^{-1}\X(\om)(\X(\om)^\top\bSi^{-1}\X(\om))^{-1}\X(\om)^\top\bSi^{-1} \\
\Pbt_j =& \ \Xbt(j)(\X(j)^\top\bSi^{-1}\X(j))^{-1}\X(j)^\top\bSi^{-1}.
\end{align*}
When the candidate model $j$ is overspecified, it follows that $R_1$, $R_2$, $R_3$, and $R_4$ are exactly $0$.
\end{thm}

Because the approximation of cAI in (\ref{eqn:cAIapp}) includes unknown parameters, we have to provide an estimator of cAI for practical use.
First, we obtain estimators of $R_1$ and $R_2$, which are polynomials of $\la$.
We define $\lah$, $\widehat{\la^2}$, and $\widehat{\la^3}$ as
\begin{align*}
\lah =& \ {n-p_j \over n-p_\om}{\sih_\om^2 \over \sih_j^2}, \quad \widehat{\la^2} = {(n-p_j)(n-p_j+2) \over (n-p_\om)(n-p_\om+2)}\left( {\sih_\om^2 \over \sih_j^2} \right)^2, \\
\widehat{\la^3} =& \ {(n-p_j)(n-p_j+2)(n-p_j+4) \over (n-p_\om)(n-p_\om+2)(n-p_\om+4)}\left( {\sih_\om^2 \over \sih_j^2} \right)^3.
\end{align*}
When $j \in \Jc_+$, it follows that
$$
{\sih_\om^2 \over \sih_j^2} \sim {\rm Be}\left( {n-p_\om \over 2}, {p_\om - p_j \over 2} \right)
$$
where ${\rm Be}(\cdot,\cdot)$ denotes the beta distribution.
This implies that $E(\lah) = E(\widehat{\la^2}) = E(\widehat{\la^3}) = 1$ for the overspecified case.
For the underspecified case, on the other hand, it follows that $E[( \sih_\om^2 / \sih_j^2 )^k] = \la^k + O(n^{-1})$ as $n \to \infty$ for $k = 1,2,3$.
This leads to an estimator of $R_2$ in the approximation of cAI given by (\ref{eqn:cAIapp}), which is given as follows:
\begin{equation}
\label{eqn:Rh2}
\widehat{R_2} = \ga \cdot {-2\widehat{\la^3} + (p_j + 4)\widehat{\la^2} - (p_j + 2) \over n}.
\end{equation}

Because $R_1$ is of order $O(n)$, we have to estimate $\la$ with higher-order accuracy in order to obtain an estimator of $R_1$ whose bias is of order $O(n^{-1})$ for the underspecified case.
To this end, we expand $E(\lah)$ up to $O(n^{-1})$ as
\begin{equation*}
E(\lah) = \la + {-2\la^3 + (p_j+2)\la^2 - p_j\la \over n} + O(n^{-2}).
\end{equation*}
Then, we obtain an estimator of $R_1$ given as
\begin{equation}
\label{eqn:Rh1}
\widehat{R_1} = \ga \cdot \left\{ \lah - {-2\widehat{\la^3} + (p_j+2)\widehat{\la^2} - p_j\lah \over n} - 1 \right\}.
\end{equation}

\begin{lem}
\label{lem:Rh1}
When the candidate model $j$ is underspecified, $\widehat{R_1}$ in {\rm (\ref{eqn:Rh1})} and $\widehat{R_2}$ in {\rm (\ref{eqn:Rh2})} are asymptotically unbiased estimators of $R_1$ and $R_2$, respectively, whose bias is of order $O(n^{-1})$.
That is,
$$
E(\widehat{R_1}) = R_1 + O(n^{-1}) \quad {\rm and} \quad E(\widehat{R_2}) = R_2 + O(n^{-1}).
$$
When the candidate model $j$ is overspecified, it follows that $E(\widehat{R_1}) = E(\widehat{R_2}) = 0$.
\end{lem}

We next consider estimation procedures for $R_3$ and $R_4$, which are complex functions of unknown parameters.
We see $R_3$ and $R_4$ as functions of $\bta_* = (\bbe_*^\top,\si^2_*)^\top$.
That is, $R_3 = R_3(\bta_*)$, $R_4 = R_4(\bta_*)$ and we substitute their unbiased estimators $\btat = (\bbet^\top,\sit^2)^\top$, which are given by
\begin{align*}
\bbet =& \ \bbeh_\om = (\X(\om)^\top\bSi^{-1}\X(\om))^{-1}\X(\om)^\top\bSi^{-1}\y, \\
\sit^2 =& \ (\y - \X(\om)\bbet)^\top \bSi^{-1} (\y - \X(\om)\bbet) / (n-p_\om).
\end{align*}
Then, the plug-in estimator of $R_4$ is
\begin{equation}
\label{eqn:R4t}
\widetilde{R_4} = R_4(\btat) = n^{-1}\{ -2\lat^3 + (p_j+4)\lat^2 \} \times \bbet^\top\B^\top\Rbt^{-1}\B\bbet / \sit^2,
\end{equation}
where $\lat = 1 / (1 + \delt)$ for
$$
\delt = \bbet^\top\X(\om)^\top(\P_\om - \P_j)\X(\om)\bbet / (n\sit^2).
$$
Because $R_3$ is of order $O(n)$, the plug-in estimator $R_3(\btat)$ has bias which is of order O(1).
We correct the bias using an analytical method based on Taylor series expansions.
We define the following estimator of $R_3$:
\begin{equation}
\label{eqn:R3t}
\widetilde{R_3} = R_3(\btat) - B_1(\btat),
\end{equation}
where
\begin{equation*}
B_1(\btat) = {\sit^2 \over 2}\cdot \tr \left[ \left. { \partial^2 R_3(\bta) \over \partial\bbe\partial\bbe^\top }\right|_{\bta = \btat} (\X(\om)^\top \bSi^{-1} \X(\om))^{-1} \right] + \left. {\partial^2 R_3(\bta) \over (\partial \si^2)^2} \right|_{\bta = \btat} \cdot {(\sit^2)^2 \over n-p_\om},
\end{equation*}
for
\begin{align*}
\left. {\partial^2 R_3(\bta) \over \partial\bbe\partial\bbe^\top} \right|_{\bta = \btat} =& \ \bbet^\top\B^\top\Rbt^{-1}\B\bbet \times \ \left\{ -{2 \C \over n(\sit^2)^2(1+\delt)^2} + {8 \C\bbet\bbet^\top\C \over n^2(\sit^2)^3(1+\delt)^3} \right\} \\
&-{4\B^\top\Rbt^{-1}\B\bbet\bbet^\top\C + 4\C\bbet\bbet^\top\B^\top\Rbt^{-1}\B \over n(\sit^2)^2(1+\delt)^2} + 2\lat \cdot \B^\top\Rbt^{-1}\B / \sit^2, \\
\left. {\partial^2 R_3(\bta) \over (\partial \si^2)^2} \right|_{\bta = \btat} =& \ \bbet^\top\B^\top\Rbt^{-1}\B\bbet \times \left\{ -{4\bbet^\top\C\bbet \over n(\sit^2)^4(1+\delt)^2} + {2(\bbet^\top\C\bbet)^2 \over n^2(\sit^2)^5(1+\delt)^3} + {2\lat \over (\sit^2)^3} \right\} \\
\C =& \ \X(\om)^\top(\P_\om - \P_j)\X(\om).
\end{align*}
Then, we have the following lemma.

\begin{lem}
\label{lem:R34t}
Both for the cases in which the candidate model $j$ is overspecified and $j$ is underspecified, $\widetilde{R_3}$ and $\widetilde{R_4}$ in {\rm (\ref{eqn:R3t})} and {\rm (\ref{eqn:R4t})} are asymptotically unbiased estimators of $R_3$ and $R_4$, whose bias is of order $O(n^{-1})$.
That is,
$$
E[\widetilde{R_3}] = R_3 + O(n^{-1}), \quad {\rm and} \quad E[\widetilde{R_4}] = R_4 + O(n^{-1}).
$$
\end{lem}

Using $\widehat{R_1}$, $\widehat{R_2}$, $\widetilde{R_3}$, and $\widetilde{R_4}$ given by (\ref{eqn:Rh1}), (\ref{eqn:Rh2}), (\ref{eqn:R3t}), and (\ref{eqn:R4t}), respectively, we construct an estimator of cAI as follows:
\begin{equation}
\label{eqn:cAIh}
\widehat{\cAI} = m\log(2\pi\sih_j^2) + \log|\Rbt| + R^* + \widehat{R_1} + \widehat{R_2} + \widetilde{R_3} + \widetilde{R_4}.
\end{equation}
Then, Theorem \ref{thm:cAIm} and Lemmas \ref{lem:Rh1}--\ref{lem:R34t} lead to the following theorem.

\begin{thm}
\label{thm:cAIh}
Both for the cases in which the candidate model $j$ is overspecified and $j$ is underspecified, $\widehat{\cAI}$ in {\rm (\ref{eqn:cAIh})} is an asymptotically unbiased estimator of $\cAI$ the bias of which is of order $O(n^{-1})$.
That is,
$$
E(\widehat{\cAI}) = \cAI + O(n^{-1}).
$$
\end{thm}

When the sample size $n$ is small, the accuracy of $\widehat{\cAI}$ does not seem to be sufficient for the overspecified model.
As the simulation study in Section \ref{subsec:simu_bias} demonstrates, $\widehat{\cAI}$ of the true model has relatively large bias, though it is important to accurately estimate cAI of the true model.
Moreover, some of the other information criteria including the cAIC of \citet{VB05}, are exact unbiased estimators of the information for an overspecified candidate model.
Thus, we should improve the estimators of $R_3$ and $R_4$ to remove the bias that is of order $O(n^{-1})$.
To this end, we adopt a parametric bootstrap method.

A bootstrap sample $\y^\dag$ is generated by
$$
\y^\dag = \X(\om)\bbet + \Z\b^\dag + \bep^\dag,
$$
where $\b^\dag$ and $\bep^\dag$ are generated according to the following distributions:
$$
\b^\dag \sim \Nc(\zero,\sit^2\G), \quad {\rm and} \quad \bep^\dag \sim \Nc(\zero,\sit^2\I_n).
$$
Then, we propose the following estimators of $R_3$ and $R_4$:
\begin{align}
\widehat{R_3} &= 2R_3(\btat) - E_\btat[R_3(\btat^\dag)] + E_\btat[B_1(\btat^\dag)] - B_1(\btat) \label{eqn:R3h} \\
\widehat{R_4} &= 2R_4(\btat) - E_\btat[R_4(\btat^\dag)], \label{eqn:R4h}
\end{align}
where $E_\btat$ denotes expectation with respect to the bootstrap distribution and $\btat^\dag $ is $\btat^\dag = ((\bbet^\dag)^\top,\sit^{2\dag})^\top$ for
\begin{align*}
\bbet^\dag =& \ (\X(\om)^\top\bSi^{-1}\X(\om))^{-1}\X(\om)\bSi^{-1}\y^\dag, \\
\sit^{2\dag} =& \ (\y^\dag - \X(\om)\bbet^\dag)^\top\bSi^{-1}(\y^\dag - \X(\om)\bbet^\dag) / (n-p_\om).
\end{align*}

%As for $R_3$, it follows from (\ref{eqn:R3bias}) that
%$$
%E_\btat [R_3(\btat^\dag)] = R_3(\btat) + B_1(\btat) + B_2(\btat) + O_p(n^{-2}).
%$$
%However, $B_1(\btat)$ has a bias with order $O(n^{-1})$, that is,
%$$
%E[B_1(\btat)] = B_1(\bta_*) + B_{11}(\bta_*) + O(n^{-2}),
%$$
%where $B_{11}(\bta_*) = O(n^{-1})$.
%Because this bias is not negligible when we want to estimate $R_3$ with third-order accuracy, we estimate the bias by bootstrap method as follows:
%$$
%\widehat{B_{11}} = E_\btat[B_1(\btat^\dag)] - B_1(\btat).
%$$
%Then, we obtain an estimator of $R_3$, which is given as
%\begin{align}
%\widehat{R_3} =& \ 2R_3(\btat) - E_\btat[R_3(\btat^\dag)] + \widehat{B_{11}} \non\\
%=& \ 2R_3(\btat) - E_\btat[R_3(\btat^\dag)] + E_\btat[B_1(\btat^\dag)] - B_1(\btat). \label{eqn:R3h}
%\end{align}

\begin{lem}
\label{lem:R34h}
Both for the cases in which the candidate model $j$ is overspecified and $j$ is underspecified, $\widehat{R_3}$ and $\widehat{R_4}$ in {\rm (\ref{eqn:R3h})} and {\rm (\ref{eqn:R4h})} are asymptotically unbiased estimators of $R_3$ and $R_4$, whose bias is of order $O(n^{-2})$.
That is,
$$
E(\widehat{R_3}) = R_3 + O(n^{-2}), \quad {\rm and} \quad E(\widehat{R_4}) = R_4 + O(n^{-2}).
$$
\end{lem}

Using $\widehat{R_1}$, $\widehat{R_2}$, $\widehat{R_3}$, and $\widehat{R_4}$ given by (\ref{eqn:Rh1}), (\ref{eqn:Rh2}), (\ref{eqn:R3h}), and (\ref{eqn:R4h}), we obtain an estimator of cAI as follows:
\begin{equation}
\label{eqn:cAIhs}
\widehat{\cAI}^\dag = m\log(2\pi\sih_j^2) + \log|\Rbt| + R^* + \widehat{R_1} + \widehat{R_2} + \widehat{R_3} + \widehat{R_4},
\end{equation}
which is higher order accurate than $\widehat{\cAI}$ in (\ref{eqn:cAIh}).
Then, we obtain the following theorem, which is proved by Theorem \ref{thm:cAIh} and Lemma \ref{lem:R34h}.

\begin{thm}
\label{thm:cAIhs}
When the candidate model $j$ is overspecified, $\widehat{\cAI}^\dag$ in {\rm (\ref{eqn:cAIhs})} is an asymptotically unbiased estimator of $\cAI$ the bias of which is of order $O(n^{-2})$, that is,
$$
E(\widehat{\cAI}^\dag) = \cAI + O(n^{-2}).
$$
When the candidate model $j$ is underspecified, $\widehat{\cAI}^\dag$ is an asymptotically unbiased estimator of $\cAI$ the bias of which is of order $O(n^{-1})$, that is,
$$
E(\widehat{\cAI}^\dag) = \cAI + O(n^{-1}).
$$
\end{thm}

Table \ref{tab:bias} shows the order of the bias of estimators for each term in $\cAI$.

\begin{table}[htbp]
  \centering
  \caption{Bias of estimators for each term in $\cAI$}
    \begin{tabular}{lccc}
    \toprule
     & $\widehat{R_1}, \ \widehat{R_2}$ & $\widetilde{R_3}, \ \widetilde{R_4}$ & $\widehat{R_3}, \ \widehat{R_4}$ \\
    \cline{2-4}
    Overspecified  & 0           & $O(n^{-1})$ & $O(n^{-2})$ \\
    Underspecified & $O(n^{-1})$ & $O(n^{-1})$ & $O(n^{-2})$ \\
    \bottomrule
    \end{tabular}%
  \label{tab:bias}
\end{table}

\section{Application to small area estimation}
\label{sec:ex}

A typical example of the covariate shift situation appears in the small area estimation problem.
The model for small area estimation supposes that the observed small area data have a finite population, which is assumed to be a realization of some super-population model with random effects, one of which is the well-known NERM proposed by \citet{BHF88}.

Let $Y_{ik}$ and $\x_{ik}(j)$ denote the value of a characteristic of interest and its $p_j$-dimensional auxiliary variable for the $k$th unit of the $i$th area for $i=1,\dots,q$ and $k=1,\dots,N_i$.
Note that $\x_{ik}(j)$ is a subvector of $\x_{ik}(\om)$, which is the vector of the explanatory variables in the full model $\om$, and we hereafter abbreviate the model index $j$ and write $\x_{ik}$ instead of $\x_{ik}(j)$, $p$ instead of $p_j$, etc.
Then, the NERM is
\begin{align}
\label{eqn:fNERM}
Y_{ik}=\x_{ik}^\top\bbe +b_i +\ep_{ik} \quad (i=1,\dots,q; \ k=1,\dots,N_i), 
\end{align}
where $\bbe$ is a $p \times 1$ vector of regression coefficients, $b_i$ is a random effect for the $i$th area, and the $b_i$s and $\ep_{ik}$s are mutually independently distributed as $b_i\sim \Nc(0,\tau^2)$ and $\ep_{ik}\sim\Nc(0,\si^2)$, respectively.
We consider the situation in which only $n_i$ values of the $Y_{ik}$s are observed through some sampling procedure.
We define the number of unobserved variables in the $i$th area by $N_i-n_i=r_i$ and let $n=n_1+\dots+n_q, r=r_1+\dots+r_q$.
Suppose, without loss of generality, the first $n_i$ elements of $\{ Y_{i1},\dots,Y_{i,N_i} \}$ are observed and denoted by $y_1,\dots,y_{i,n_i}$, and that $Y_{i,n_i+1},\dots,Y_{i,N_i}$ are unobserved. Then, the model for observed data is defined as
\begin{equation}
\label{eqn:oNERM}
y_{ik}=\x_{ik}^\top\bbe +b_i +\ep_{ik} \quad (i=1,\dots,q; \ k=1,\dots,n_i),
\end{equation}
which corresponds to (\ref{eqn:omodel}) with $\y=(\y_1^\top,\ldots,\y_q^\top)^\top$ for $\y_i=(y_{i1},\dots,y_{i,n_i})^\top$, $\X=(\X_1^\top,\dots,\X_q^\top)^\top$ for $\X_i=(\x_{i1},\dots,\x_{i,n_i})^\top$, $\Z=\diag(\Z_1,\dots,\Z_q)$ for $\Z_i=\one_{n_i}$, $\G=\psi\I_q$ and $\R=\I_n$, where $\one_{n_i}$ denotes an $n_i \times 1$ vector of 1s and $\psi=\tau^2/\si^2$.
%Note that $q=k$.
In the derivation of our proposed criteria, we assume that the covariance matrix of $\b$ is $\si^2\G$ for a known matrix $\G$.
However, in the NERM, $\G$ includes the parameter $\psi$, which is usually unknown and has to be estimated.
In this case, we propose that $\G$ in the bias correction should be replaced with its plug-in estimator $\G(\psih)$ for some consistent estimator $\psih$.
We recommend that $\psi$ should be estimated based on the full model.

We consider two types of models for prediction.
The first can be used in the situation in which all $\x_{ik}$s are available.
Then, the model for prediction, which we call the ``unit-level model for prediction'', is defined by
\begin{align}
\label{eqn:p1NERM}
Y_{ik} = \x_{ik}^\top\bbe +b_i +\ep_{ik} \quad (i=1,\dots,q; \ k=n_i+1,\dots,N_i), 
\end{align}
which corresponds to (\ref{eqn:pmodel}) with $\ybt=(\ybt_1^\top,\dots,\ybt_q^\top)^\top$ for $\ybt_i=(Y_{i,n_i+1},\dots,Y_{i,N_i})^\top$, $\Xbt=(\Xbt_1^\top,\dots,\Xbt_q^\top)^\top$ for $\Xbt_i=(\x_{i,n_i+1},\dots,\x_{i,N_i}^\top)^\top$, $\Zbt=\diag(\Zbt_1,\dots,\Zbt_q)$ for $\Zbt_i=\one_{r_i}$, $\Rbt=\I_r$.
Note that $m=r$.

In the problem of small area estimation, we often encounter the situation in which all $\x_{ik}$s are not observed but the area mean $\bar{\x}_i=N_i^{-1}\sum_{k=1}^{N_i}\x_{ik}$ is known and we are interested in predicting $\Yo_i$, which is the mean of the finite population $\{ Y_{i1},\dots,Y_{i,N_i} \}$, by using the value of $\bar{\x}_i$.
Then, the second type of model for prediction, which we call the ``area-level model for prediction,'' is defined as
\begin{equation}
\label{eqn:p2NERM}
\Yo_{i(u)}=\bar{\x}_{i(u)}^\top\bbe +b_i +\bar{\ep}_{i(u)} \quad (i=1,\dots,q), 
\end{equation}
where $\Yo_{i(u)}=r_i^{-1}\sum_{k=n_i+1}^{N_i}Y_{ik}$, the mean of unobserved variables, $\bar{\x}_{i(u)}=r_i^{-1}\sum_{k=n_i+1}^{N_i}\x_{ik}$, calculated from $\bar{\x}_{i}$ and $(\x_{i1},\ldots,\x_{in_i})$, and $\bar{\ep}_{i(u)}=r_i^{-1}\sum_{k=n_i+1}^{N_i}\ep_{ik}$ is distributed as $\Nc(0,\sigma^2/r_i)$.
The model (\ref{eqn:p2NERM}) corresponds to (\ref{eqn:pmodel}) with $\ybt=(\Yo_{1(u)},\dots,\Yo_{q(u)})^\top$, $\Xbt=(\bar{\x}_{1(u)},\dots,\bar{\x}_{q(u)})^\top$, $\Zbt=\I_q$ and $\Rbt=\diag(\Rt_1,\dots,\Rt_q)$ for $\Rt_i=1/r_i$.
Note that $m=q$.

After selecting explanatory variables with our proposed criteria, we predict $\Yo_{i(u)}$ using the empirical best linear unbiased predictor $\widehat{\Yo}_{i(u)}=\bar{\x}_{i(u)}^\top\bbeh +\bh_i$ and obtain a predictor of the finite population mean $\Yo_i$, which is given as
\begin{equation}
\label{eqn:fp}
\widehat{\Yo}_{i}=\frac{1}{N_i} \left\{ \sum_{k=1}^{n_i}y_{ik} +r_i\widehat{\Yo}_{i(u)} \right\}.
\end{equation}
Thus, covariate shift appears in standard models for small area estimation and the proposed criterion is important and useful in such situations.

\section{Numerical studies}
\label{sec:simu}

\subsection{Measuring the bias of estimating the true cAI by the criteria}
\label{subsec:simu_bias}

In this subsection, we compare the performance of the criteria by measuring the bias of estimating the cAI.
We consider a class of nested candidate models $j_\al = \{ 1,\dots,\al \}$ for $\al = 1,\dots,p_\om$ where $p_\om = 7$.
The true model for observed data is the NERM in (\ref{eqn:oNERM}) with $\si^2 = \tau^2 = 1$ and $n_i = r_i = 3$ for $i=1,\dots,q$.
We consider the unit-level model for prediction (\ref{eqn:p1NERM}) for the first experiment and the area-level model for prediction (\ref{eqn:p2NERM}) for the second experiment.
The explanatory variables in the full model, $\x_{ik}(\om)$'s ($i=1,\dots,q; \ k=1,\dots,N_i$), are independently generated by $\Nc(\zero,\bSi_\x)$, where $\bSi_\x = 0.9\I_{p_\om} + 0.1\J_{p_\om}$ for $\J_{p_\om} = \one_{p_\om}\one_{p_\om}^\top$.
The true coefficient vector $\bbe_*$ is $\bbe_* = (\be_1,\dots,\be_{p_*},0,0)$ for $p_* = 5$ and $\be_l$s ($1\leq l \leq p_*$) are generated by $\be_l = 2 \times ((-1)^l / (l+0.7)) \times U(1,2)$ for a uniform random variable $U(1,2)$ on the interval $(1,2)$.
The values of the explanatory variables $\x_{ik}$s and the vector of regression coefficients $\bbe_*$ are fixed throughout the simulations.

For comparison, we consider the exact unbiased estimator, which is derived under the assumption that the candidate model is overspecified, given by
\begin{equation}
\label{eqn:CScAIC}
\widehat{\cAI}_{\rm u} = m\log(2\pi\sih_j^2) + \log|\Rbt| + (\y - \X(j)\bbeh_j - \Z\bbh_j)^\top \R^{-1} (\y - \X(j)\bbeh_j - \Z\bbh_j) / \sih_j^2 + \De_{\rm CS},
\end{equation}
where the bias correction term $\De_{\rm CS}$ is
\begin{align*}
\De_{\rm CS} =& \ {n \over n-p_j-2}\left\{ \tr[\Rbt^{-1}\bLa] + \tr[\Rbt^{-1}\A(\X^\top\bSi^{-1}\X)^{-1}\A^\top] \right\} \\
& + {n \over n-p_j} \left\{ -\tr[\R\bSi^{-1}] + \tr[\R\P_j] \right\}.
\end{align*}

\begin{table}[htbp]
  \centering
  \caption{Relative bias of estimating cAI by $\widehat{\cAI}_{\rm u}$, $\widehat{\cAI}$ and $\widehat{\cAI}^\dag$ when using the unit-level model for prediction.}
    \begin{tabular}{rrrrrr}
    \toprule
    \multicolumn{1}{c}{} & \multicolumn{1}{c}{True value} & \multicolumn{1}{c}{} & \multicolumn{3}{c}{Relative bias} \\
\cline{2-2}\cline{4-6}
    \multicolumn{1}{c}{Model} & \multicolumn{1}{c}{cAI} & \multicolumn{1}{c}{} & \multicolumn{1}{c}{$\widehat{\cAI}_{\rm u}$} & \multicolumn{1}{c}{$\widehat{\cAI}$} & \multicolumn{1}{c}{$\widehat{\cAI}^\dag$} \\
    \midrule
    \multicolumn{6}{c}{Pattern (a): $q=10$} \\
    \midrule
   $j_1$   & 206.38 &       & -33.439 & -0.33371 & -0.080972 \\
   $j_2$   & 152.81 &       & -18.840 & -0.2414 & -0.16414 \\
   $j_3$   & 140.79 &       & -18.556 & -0.23026 & -0.46789 \\
   $j_4$   & 132.61 &       & -11.451 & 0.26445 & -0.19514 \\
   $j_5$   & 116.51 &       & -0.0019291 & 1.5979 & 0.49514 \\
   $j_6$   & 122.46 &       & -0.050686 & 0.77756 & 0.15429 \\
   $j_7$   & 128.88 &       & 0.0086256 & 0.09468 & 0.09468 \\
    \midrule
    \multicolumn{6}{c}{Pattern (b): $q=15$} \\
    \midrule
   $j_1$  & 233.35 &       & -0.22534 & 0.11255 & 0.1159 \\
   $j_2$  & 189.54 &       & 9.4310 & 0.24145 & 0.21667 \\
   $j_3$  & 177.28 &       & 14.197 & 0.42246 & 0.37347 \\
   $j_4$  & 163.62 &       & -0.76563 & 0.32597 & 0.02934 \\
   $j_5$  & 152.94 &       & 0.13627 & 0.8566 & 0.25115 \\
   $j_6$  & 156.73 &       & 0.068668 & 0.53817 & 0.15002 \\
   $j_7$  & 161.65 &       & 0.083897 & 0.015869 & 0.015869 \\
    \midrule
    \multicolumn{6}{c}{Pattern (c): $q=20$} \\
    \midrule
   $j_1$  & 299.12 &       & 4.3775 & 0.084838 & 0.084654 \\
   $j_2$  & 252.60 &       & 6.1677 & 0.24072 & 0.23581 \\
   $j_3$  & 250.27 &       & -5.1825 & -0.016634 & 0.010911 \\
   $j_4$  & 208.25 &       & 2.6115 & 0.36013 & 0.23929 \\
   $j_5$  & 197.52 &       & 0.25682 & 0.53321 & 0.26101 \\
   $j_6$  & 200.38 &       & 0.24977 & 0.40855 & 0.25647 \\
   $j_7$  & 203.57 &       & 0.22713 & 0.21272 & 0.21272 \\
    \bottomrule
    \end{tabular}%
  \label{tab:biasu}
\end{table}

\begin{table}[htbp]
  \centering
  \caption{Relative bias of estimating cAI by $\widehat{\cAI}_{\rm u}$, $\widehat{\cAI}$ and $\widehat{\cAI}^\dag$ when using the area-level model for prediction.}
    \begin{tabular}{rrrrrr}
    \toprule
    \multicolumn{1}{c}{} & \multicolumn{1}{c}{True value} & \multicolumn{1}{c}{} & \multicolumn{3}{c}{Relative bias} \\
    \cline{2-2}\cline{4-6}
    \multicolumn{1}{c}{Model} & \multicolumn{1}{c}{cAI} & \multicolumn{1}{c}{} & \multicolumn{1}{c}{$\widehat{\cAI}_{\rm u}$} & \multicolumn{1}{c}{$\widehat{\cAI}$} & \multicolumn{1}{c}{$\widehat{\cAI}^\dag$} \\
    \midrule
    \multicolumn{6}{c}{Pattern (a): $q=10$} \\
    \midrule
   $j_1$  & 61.095 &       & -10.429 & -0.27157 & -0.21359 \\
   $j_2$  & 46.635 &       & 5.4222 & 0.40292 & -0.055653 \\
   $j_3$  & 50.744 &       & -10.285 & 0.36735 & -0.23857 \\
   $j_4$  & 47.323 &       & -0.77484 & 1.3412 & 0.36198 \\
   $j_5$  & 45.735 &       & -0.21108 & 2.2751 & 0.60073 \\
   $j_6$  & 49.452 &       & -0.32466 & 0.82334 & 0.017969 \\
   $j_7$  & 52.805 &       & -0.29278 & -0.082743 & -0.082743 \\
    \midrule
    \multicolumn{6}{c}{Pattern (b): $q=15$} \\
    \midrule
   $j_1$  & 95.393 &       & -5.4716 & 0.12331 & 0.13930 \\
   $j_2$  & 70.056 &       & 17.521 & 0.39905 & 0.36599 \\
   $j_3$  & 66.412 &       & 21.039 & 0.59611 & 0.53872 \\
   $j_4$  & 61.310 &       & 5.1740 & 0.58453 & 0.10853 \\
   $j_5$  & 60.532 &       & 0.23723 & 1.0853 & 0.25515 \\
   $j_6$  & 63.109 &       & 0.13276 & 0.50306 & 0.096935 \\
   $j_7$  & 65.379 &       & 0.16648 & -0.0017143 & -0.0017143 \\
    \midrule
    \multicolumn{6}{c}{Pattern (c): $q=20$} \\
    \midrule
   $j_1$  & 98.841 &       & 21.017 & 0.18161 & 0.17565 \\
   $j_2$  & 94.83 &       & 10.059 & 0.31607 & 0.30894 \\
   $j_3$  & 88.346 &       & 5.6464 & 0.13835 & 0.11302 \\
   $j_4$  & 78.383 &       & 7.8734 & 0.46942 & 0.27593 \\
   $j_5$  & 77.794 &       & 0.18930 & 0.54202 & 0.17009 \\
   $j_6$  & 79.277 &       & 0.18908 & 0.41365 & 0.18534 \\
   $j_7$  & 81.216 &       & 0.15395 & 0.11783 & 0.11783 \\
    \bottomrule
    \end{tabular}%
  \label{tab:biasa}
\end{table}

Tables \ref{tab:biasu} and \ref{tab:biasa} report the true values of the cAI and the relative bias of estimating the cAI by the criteria $\widehat{\cAI}$ in (\ref{eqn:cAIh}), $\widehat{\cAI}^\dag$ in (\ref{eqn:cAIhs}), and $\widehat{\cAI}_{\rm u}$ in (\ref{eqn:CScAIC}), for the experiment using the unit-level model for prediction and for that using the area-level model for prediction, respectively.
We consider cases in which the number of areas $q = 10, \ 15, \ 20$.
The true values of cAI in each candidate model are calculated based on (\ref{eqn:CScAI}) with 10,000 Monte Carlo iterations.
The relative bias of estimating the cAI by the criteria is defined as
$$
100 \times {E[\IC] - \cAI \over \cAI},
$$
where $\IC = \widehat{\cAI}, \ \widehat{\cAI}^\dag$, $\widehat{\cAI}_{\rm u}$, and expectation is computed based on 1,000 replications.
The bootstrap sample size is 1,000 for obtaining $\widehat{\cAI}^\dag$.
From the tables, we observe the following: first, $\widehat{\cAI}_{\rm u}$ has large bias for underspecified models, that is, $j_1, \ j_2, \ j_3$, and $j_4$, while the modified estimators of the cAI, $\widehat{\cAI}$ and $\widehat{\cAI}^\dag$, have smaller bias for both overspecified and underspecified models; second, $\widehat{\cAI}^\dag$ can estimate the cAI better than $\widehat{\cAI}$ can for the case of small sample size because the bias of $\widehat{\cAI}^\dag$ is of order $O(n^{-2})$ while that of $\widehat{\cAI}$ is $O(n^{-1})$.
In particular, the improvement is remarkable for the true model $j_5$, which is important for variable selection.
However, the relative bias of $\widehat{\cAI}$ becomes smaller as the sample size becomes larger and the difference in performance between $\widehat{\cAI}$ and $\widehat{\cAI}^\dag$ is less significant.

\subsection{Design-based simulation based on real data}

In this subsection, we investigate the numerical performance of the small area estimation problem described in Section \ref{sec:ex}.
We conduct a design-based simulation based on a real dataset.
We use the posted land price data along the Keikyu train line, which connects the suburbs in Kanagawa prefecture to the Tokyo metropolitan area.
This dataset was also used by \citet{KK14}, who studied modification of the cAIC.

We analyze the land price data in 2001 with 47 stations that we consider as small areas, and let $q=47$.
In the original sample, there are $n_i$ sampled land spots for the $i$th area, and the total sample size is $n = \sum_{i=1}^q n_i = 189$.
We generate a synthetic population of size $N = 1,000$ by resampling with replacement from the original dataset using selection probabilities inversely proportional to sample weights.
This method of making a synthetic population was also used by \citet{CSCT12}.
Then, we select 200 independent random samples, each of size $n=189$, from the fixed synthetic population by sampling from each area based on simple random sampling without replacement and with the sample size of each area equal to that of the original dataset $n_i$.

The characteristic of interest is the land price (Yen in hundreds of thousands) per ${\rm m}^2$ of the $k$th spot in the $i$th area, denoted by $P_{ik}$, and the target is the mean of the land price in each area $\Po_i = N_i^{-1}\sum_{k=1}^{N_i} P_{ik}$ for $i = 1,\dots,q$, where $N_i$ is the size of the $i$th area (subpopulation).
As discussed in Section \ref{sec:ex}, we adopt model-based estimation of the finite subpopulation mean $\Po_i$ by using NERM.
For selecting the explanatory variables in NERM, we use our proposed criterion $\widehat{\cAI}$ in (\ref{eqn:cAIh}) for comparison with the conventional cAIC by \citet{VB05}.
However, because the land price data are right-skewed, we undertake log-transformation, namely, $Y_{ik} = \log(P_{ik})$, and fit $Y_{ik}$ with NERM in (\ref{eqn:fNERM}).

The dataset includes the following auxiliary variables.
${\rm FAR}_{ik}$ denotes the floor-area ratio of spot $k$ in the $i$th area, ${\rm TRN}_i$ is the time it takes by train from station $i$ to Tokyo station around 9:00 AM, ${\rm DST}_{ik}$ is the geographical distance from spot $k$ to nearby station $i$ and ${\rm FOOT}_{ik}$ denotes the time it takes on foot from spot $k$ to nearby station $i$.
As the candidate explanatory variables, we consider 7 variables ${\rm FAR}_{ik}$, ${\rm TRN}_i$, ${\rm TRN}_i^2$, ${\rm DST}_{ik}$, ${\rm DST}_{ik}^2$, ${\rm FOOT}_{ik}$, and ${\rm FOOT}_{ik}^2$, which are denoted by $x_1,\dots,x_7$, and $x_0 = 1$ denotes a constant term.
Using the criteria, we select the best combination of these variables.

In order to apply our criteria to the NERM, we have to estimate $\psi = \tau^2 / \si^2$, which is included in $\G = \G(\psi)$.
We estimate $\psi$ by $\psih = \tah^{\rm 2(PR)}_\om / \sih^{\rm 2(PR)}_\om$, where $\tah^{\rm 2(PR)}_\om$ and $\sih^{\rm 2(PR)}_\om$ denote the unbiased estimators by \citet{PR90} of $\tau^2$ and $\si^2$ based on the full model, respectively.

Based on the best model selected by the criteria, we obtain a predictor $\widehat{\Po}_i$ of the finite subpopulation mean of the land price in the $i$th area. 
However, log-transformed variable $Y_{ik}$ is used in the NERM, and thus, we have to modify the predictor (\ref{eqn:fp}).
The best predictor of out-of-sample $P_{ik} \ (i=1,\dots,q; \ k=n_i+1,\dots,N_i)$ is the conditional expectation given the data $\y$, namely, $\widehat{P}_{ik}(\y, \bth) = E[ P_{ik} \mid \y ] = E[ \exp(Y_{ik}) \mid \y ]$, where $\bth = (\bbe^\top,\ta^2,\si^2)^\top$.
Because the conditional mean and variance of $Y_{ik}$ given $\y$, denoted by $\mu_{ik}$ and $V_i$, are
$$
\mu_{ik} = \x_{ik}^\top\bbe + {\ta^2 \over \si^2 + n_i\ta^2}\sum_{k'=1}^{n_i}(y_{ik'} - \x_{ik'}^\top\bbe), \quad V_i = \si^2 + {\ta^2\si^2 \over \si^2 + n_i\ta^2},
$$
it follows that
$$
\widehat{P}_{ik}(\y,\bth) = \exp( \mu_{ik} + V_i/2 ),
$$
Substituting $\bthh$ with $\bth$, we obtain the empirical best predictor (EBP):
$$
\widehat{P}_{ik}^{\rm EBP} = \widehat{P}_{ik}(\y,\bthh),
$$
where $\bthh$ is some estimator of $\bth$.
Then, we use the following predictor of $\Po_i$:
$$
\widehat{\Po}_i = {1 \over N_i} \left\{ \sum_{k=1}^{n_i}P_{ik} + \sum_{k=n_i+1}^{N_i}\widehat{P}_{ik}^{\rm EBP} \right\}.
$$
As $\bthh$, we use unbiased estimators of $\ta^2$ and $\si^2$ proposed by \citet{PR90} and the GLS estimator of $\bbe$.

\begin{figure}
\begin{center}
\caption{MSE ratios of cAIC by \citet{VB05} to $\widehat{\cAI}$ based on area-level model for prediction (solid line), to $\widehat{\cAI}_{\rm u}$ (dashed line) and to fence method by \citet{JRGN08} (dotdash line).}
\includegraphics[width=\textwidth]{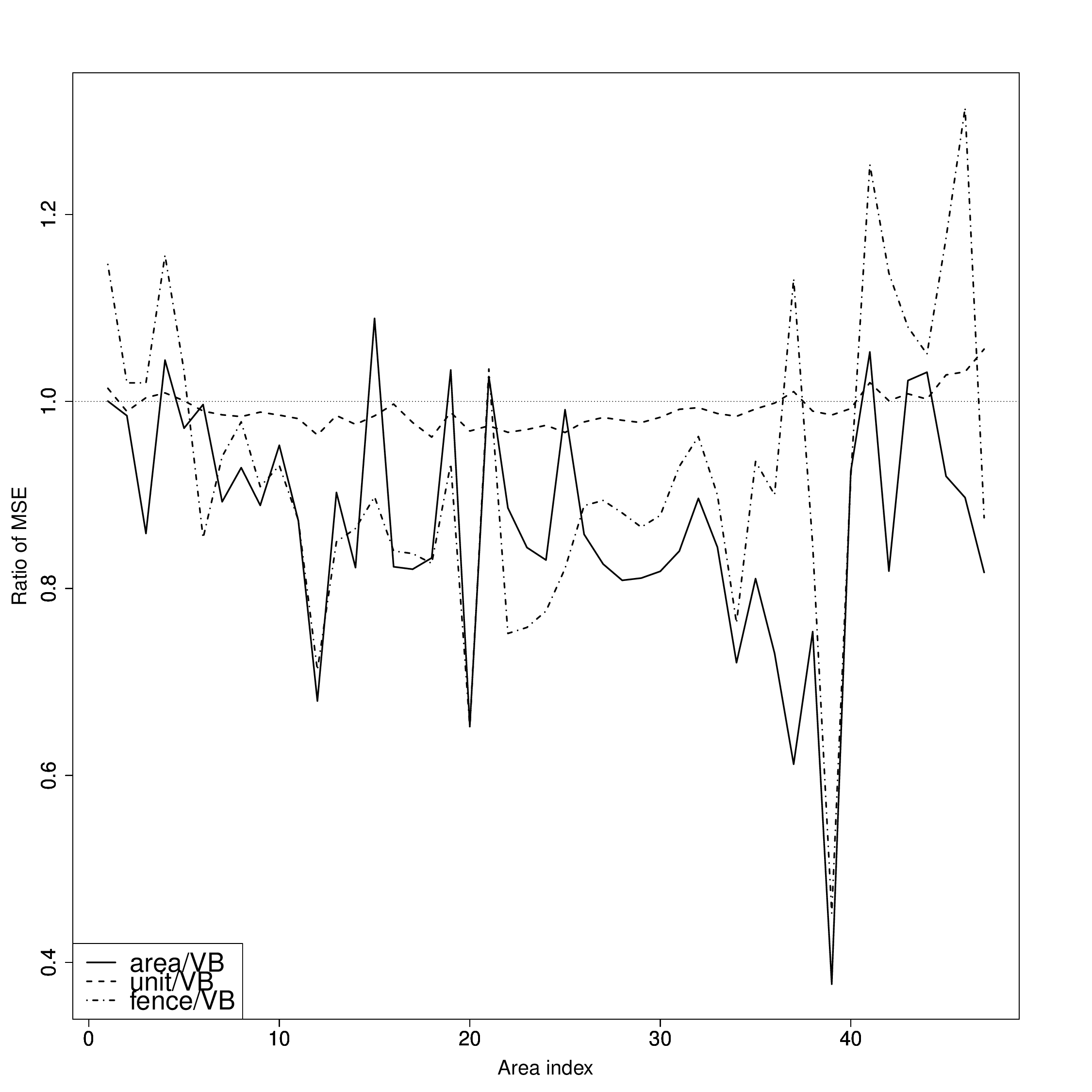}
\end{center}
\label{fig:MSE}
\end{figure}

We measure the performance of this design-based simulation by calculating the mean squared error (MSE) of the predictor,
$$
{\rm MSE}_i = {1 \over 200} \sum_{t=1}^{200} ( \widehat{\Po}_{i}^{(t)} - \Po_i^{(t)} )^2,
$$
where $\Po_i^{(t)}$ and $\widehat{\Po}^{(t)}$ are the $i$th finite subpopulation mean and its predictor of the $t$th sampled data.
We construct $\widehat{\cAI}$s based on the unit-level model for prediction (\ref{eqn:p1NERM}), denoted by $\widehat{\cAI}_{\rm u}$, and based on the area-level model for prediction (\ref{eqn:p2NERM}), denoted by $\widehat{\cAI}_{\rm a}$, and let ${\rm MSE}^{\rm u}_i$ and ${\rm MSE}^{\rm a}_i$ denote the corresponding MSEs.
To compare the performance, we compute ratios of MSEs as follows:
$$
{\MSE^{\rm u}_i \over \MSE^{\rm vb}_i}, \quad {\MSE^{\rm a}_i \over \MSE^{\rm vb}_i}, \quad {\rm and} \quad { \MSE^{\rm fence}_i \over \MSE^{\rm vb}_i }
$$
where $\MSE_i^{\rm vb}$ and $\MSE_i^{\rm fence}$ are the MSEs of the predictor based on the cAIC of \citet{VB05} and the fence method proposed by \citet{JRGN08}, respectively.
To implement the fence method, we use the \texttt{fence.lmer()} function in the R package `fence'.
Figure \ref{fig:MSE} shows the results.
Although the performance of $\widehat{\cAI}_{\rm u}$ is similar to the cAIC of \citet{VB05}, $\widehat{\cAI}_{\rm a}$ has much better performance in most areas.
The performance of $\widehat{\cAI}_{\rm a}$ and that of the fence method are similar, both of which tend to select parsimonious models, though $\widehat{\cAI}_{\rm a}$ performs better than the fence method in most areas.
It is valuable to point out that the MSE of the predictor of the finite subpopulation mean can be improved using our proposed criteria, which motivates us to use them for variable selection in the small area estimation problem.

Finally, we apply our method to the original land price dataset.
Table \ref{tab:SV} shows the variables selected by several methods.
We can see from the table that cAIC by \citet{VB05} and $\widehat{\cAI}_{\rm u}$ select the same large model and that $\widehat{\cAI}_{\rm a}$ and the fence method select more parsimonious models.
The estimates of the variance parameters for the model selected by $\widehat{\cAI}$ based on the area-level model are $\tah^2 = 0.0103299$ and $\sih^2 = 0.0230414$.
This implies that our $\widehat{\cAI}$ method produces reasonable results for models which do not have a common variance parameter, though we assume this when deriving the criteria.

\begin{table}[htbp]
  \centering
  \caption{Variables selected by each method. VB, Unit, Area and fence denote cAIC by \citet{VB05}, $\widehat{\cAI}$ based on unit-level model, $\widehat{\cAI}$ based on area-level model and fence method, respectively.}
    \begin{tabular}{ll}
    \toprule
    VB  & $x_0, x_1, x_2, x_3, x_6, x_7$ \\
    Unit & $x_0, x_1, x_2, x_3, x_6, x_7$ \\
    Area & $x_0, x_1, x_2, x_3$ \\
    fence & $x_0, x_1, x_2$ \\
    \bottomrule
    \end{tabular}%
  \label{tab:SV}
\end{table}

\section{Discussion}
\label{sec:dis}

We have redefined the conditional Akaike information (cAI) under the covariate shift situation.
We have derived an asymptotically unbiased estimator of the cAI without assuming that the candidate model is overspecified, which can work as a variable selection criterion.
The resulting criterion is applied to estimating the finite population mean in the context of the small area estimation problem.
Numerical studies demonstrate that our proposed procedure performs well.

In the application to estimating the finite population mean, we have considered two models for prediction, one is the unit-level model and the other is the area-level model.
We have found that the area-level model performs well in the design-based simulation though the theoretical reason is not clear.
Examining more detailed theoretical properties of our approach would be an interesting problem, which will be left as future work.

Note that we assumed the same variance component for both the random effects and the error terms when we derived the criterion.
Although numerical studies showed that this assumption is not very restrictive in practice, relaxation of this restriction will also be an important direction of future work.

\medskip
\bigskip
\noindent
{\bf Acknowledgments}

%The authors are grateful to the Editors and the anonymous referees for their valuable comments and helpful suggestions.
We are grateful to Professor J.N.K. Rao for his helpful comment and for holding a seminar at Statistics Canada during our stay in Ottawa.
This research was supported by Grant-in-Aid for Scientific Research from the Japan Society for the Promotion of Science, Grant Numbers 16K17101, 16H07406, 15H01943 and 26330036.
%The computational results were obtained using Ox version 6.21.
%Ox codes are available as the Supplementary Materials.

\appendix
\section{Appendix}
\subsection{Proof of Theorem \ref{thm:cAIm}}

Because $\bbeh_j$ and $\sih_j^2$ are mutually independent, cAI in (\ref{eqn:CScAI}) can be rewritten as
$$
\cAI = E[m\log(2\pi\sih_j^2)] + \log|\Rbt| + n\cdot E\big[ (n\sih_j^2 / \si_*^2)^{-1} \big] \big\{ \tr(\Rbt^{-1}\bLa) + E[\a^\top\Rbt^{-1}\a / \si_*^2] \big\},
$$
For the overspecified case, we can easily evaluate cAI, noting that $n\sih_j^2 / \si_*^2$ follows a chi-squared distribution with $n-p_j$ degrees of freedom and is independent of $\bbeh_j$.
Thus, it suffices to show that cAI is evaluated as (\ref{eqn:cAIapp}) for the underspecified case.

From (B.4) in \citet{KK14}, we can evaluate $E[(n\sih_j^2 / \si_*^2)^{-1}]$ as follows:
\begin{equation}
\label{eqn:se}
E[(n\sih_j^2 / \si_*^2)^{-1}] = {\la \over n} \left\{ 1 + {-2\la^2 + (p_j+4)\la \over n} \right\} + O(n^{-3}).
\end{equation}
We next evaluate $E[\a^\top\Rbt^{-1}\a / \si_*^2]$.
Let $\u = \y - \X(\om)\bbe_* \sim \Nc_n(\zero,\si_*^2\bSi)$.
Then, we can rewrite $\Xbt(j)\bbeh_j - \Xbt(\om)\bbe_*$ in $\a$ as
\begin{align*}
\Xbt(j)\bbeh_j - \Xbt(\om)\bbe_* =& \ \Pbt_j(\X(\om)\bbe_* + \u) - \Xbt(\om)\bbe_* \\
=& \ (\Pbt_j\X(\om) - \Xbt(\om))\bbe_* + \Pbt_j\u.
\end{align*}
Next, we can rewrite $\X(j)\bbeh_j - \X(\om)\bbe_*$ in $\a$ as
\begin{align*}
\X(j)\bbeh_j - \X(\om)\bbe_*
=& \ \bSi \{ \P_j( \X(\om)\bbe_* + \u ) - \bSi^{-1}\X(\om)\bbe_* \} \\
=& \ -\bSi(\P_\om - \P_j)\X(\om)\bbe_* + \bSi\P_j\u.
\end{align*}
Then, we obtain
\begin{equation*}
\a = \B\bbe_* + (\Pbt_j - \Zbt\G\Z^\top\P_j)\u.
\end{equation*}
Moreover, it follows that
\begin{align*}
\Pbt_j - \Zbt\G\Z^\top\P_j =& \ (\Xbt(j) - \Zbt\G\Z^\top\bSi^{-1}\X(j))(\X(j)^\top\bSi^{-1}\X(j))^{-1}\X(j)^\top\bSi^{-1} \\
=& \ \A(\X(j)^\top\bSi^{-1}\X(j))^{-1}\X(j)^\top\bSi^{-1}
\end{align*}
Thus, $E[\a^\top\Rbt^{-1}\a / \si_*^2]$ can be evaluated as
\begin{equation}
\label{eqn:aRa}
E[\a^\top\Rbt^{-1}\a / \si_*^2] = \tr[\Rbt^{-1}\A(\X(j)^\top\bSi^{-1}\X(j))^{-1}\A^\top] + \bbe_*^\top\B^\top\Rbt^{-1}\B\bbe_* / \si_*^2.
\end{equation}
It follows from (\ref{eqn:se}) and (\ref{eqn:aRa}) that
\begin{align*}
& \ n\cdot E[(n\sih_j^2/\si_*^2)^{-1}]\{ \tr(\Rbt^{-1}\bLa) + E[\a^\top\Rbt^{-1}\a / \si_*^2] \} \\
=& \ \left\{ \la + {-2\la^3 + (p_j+4)\la^2 \over n} \right\} \times (\ga + \bbe_*^\top\B^\top\Rbt^{-1}\B\bbe_* / \si_*^2) + O(n^{-1}),
\end{align*}
which indicates that cAI can be approximated as (\ref{eqn:cAIapp}).
\hfill$\Box$

\subsection{Proof of Lemma \ref{lem:Rh1}}
This lemma can be proved using Appendix C and Appendix D of \citet{KK14}.

%\subsection{Proof of Lemma \ref{lem:der}}
%
%First, note that
%$$
%R_3(\bta_*) = \la \cdot \bbe_*^\top\B^\top\Rbt^{-1}\B\bbe_* / \si_*^2,
%$$
%where $\la = 1 / (1+\de)$ for
%$$
%\de = \bbe_*^\top\X(\om)^\top(\P_\om - \P_j)\X(\om)\bbe_* / (n\si_*^2).
%$$
%Then, we observe that
%$$
%{\partial R_3(\bta_*) \over \partial \bta_*} = {\partial \la \over \partial \de}\cdot {\partial \de \over \partial \bta_*}
%$$
%and that $\partial \la / \partial \de = -(1+\de)^{-2}$.
%After some calculations, we obtain Lemma \ref{lem:der}.
%\hfill$\Box$

\subsection{Proof of Lemma \ref{lem:R34t}}

First, note that $R_3(\btat)$ is expanded as
\begin{equation*}
R_3(\btat) = R_3(\bta_*) + {\partial R_3(\bta_*) \over \partial \bta_*^\top}(\btah - \bta_*) + {1 \over 2}(\btah - \bta_*)^\top {\partial^2 R_3(\bta_*) \over \partial \bta_* \partial \bta_*^\top}(\btah - \bta_*) + o(1).
\end{equation*}
Given that $\btat$ is an unbiased estimator of $\bta_*$ and $\bbet$ and $\sit^2$ are mutually independent, it follows that
\begin{equation*}
E[R_3(\btat)] = R_3(\bta_*) + B_1(\bta_*) + O(n^{-1}),
\end{equation*}
where
\begin{equation}
\label{eqn:B1}
\begin{split}
B_1(\bta_*) &= {1 \over 2} \cdot \tr \left[ {\partial^2 R_3(\bta_*) \over \partial\bbe_* \partial\bbe_*^\top}E[(\bbet - \bbe_*)(\bbet - \bbe_*)^\top] \right] + {1\over 2} {\partial^2 R_3(\bta_*) \over (\partial \si_*^2)^2}E[(\sit^2 - \si_*^2)^2] \\
&= {\si_*^2 \over 2} \cdot \tr \left[ {\partial^2 R_3(\bta_*) \over \partial\bbe_* \partial \bbe_*^\top} (\X(\om)^\top \bSi^{-1} \X(\om))^{-1} \right] + {\partial^2 R_3(\bta_*) \over (\partial \si_*^2)^2}{(\si_*^2)^2 \over n-p_\om}.
\end{split}
\end{equation}
Because $B_1(\bta_*) = O(1)$, we obtain
$$
E[\widetilde{R_3}] = E[ R_3(\btat) - B_1(\btat) ] = R_3(\bta_*) + O(n^{-1}).
$$
As for $\widetilde{R_4}$, because $R_4 = O(1)$, it follows that $E[\widetilde{R_4}] = E[R_4(\btat)] = R_4(\bta_*) + O(n^{-1})$.
\hfill$\Box$

%\bigskip
%First, note that $E[R_3(\btat)]$ is expanded as
%$$
%E[R_3(\btat)] = R_3(\bta_*) + B_1(\bta_*) + O(n^{-1}),
%$$
%where $B_1(\bta_*)$ is given as (\ref{eqn:B1}).
%Because $B_1(\bta_*) = O(1)$, it follows that $B_1(\btat) = B_1(\bta_*) + O(n^{-1})$, which shows that
%$$
%E[\widetilde{\widetilde{R_3}}] = R_3 + O(n^{-1}).
%$$
%In the same way, we obtain $E[\widetilde{R_4}] = E[R_4(\btat)] = R_4(\bta_*) + O(n^{-1})$.

\subsection{Proof of Lemma \ref{lem:R34h}}

Because $R_3(\bta_*) = O(n)$, we can see that $E[R_3(\btat)] = R_3(\bta_*) + B_1(\bta_*) + B_2(\bta_*) + O(n^{-2})$, where $B_1(\bta_*)$ is given by (\ref{eqn:B1}) and $B_2(\bta_*) = O(n^{-1})$.
Thus, it follows that
\begin{align*}
E\left[ 2R_3(\btat) - E_\btat[R_3(\btat^\dag)] \right] =& \ 2\left\{ R_3(\bta_*) + B_1(\bta_*) + B_2(\bta_*) \right\} - E\left[ R_3(\btat) + B_1(\btat) + B_2(\btat) \right] + O(n^{-2}) \\
=& \ R_3(\bta_*) + B_1(\bta_*) + B_2(\bta_*) - E\left[ B_1(\btat) + B_2(\btat) \right] + O(n^{-2}).
\end{align*}
Moreover, because $E[B_1(\btat)] = B_1(\bta_*) + B_{11}(\bta_*) + O(n^{-2})$ and $E[B_2(\btat)] = B_2(\bta_*) + O(n^{-2})$, the equation above can be rewritten as
\begin{equation}
\label{eqn:bias1}
E\left[ 2R_3(\btat) - E_\btat[R_3(\btat^\dag)] \right] = R_3(\bta_*) - B_{11}(\bta_*) + O(n^{-2}).
\end{equation}
Next, it is observed that
\begin{align}
E\left[ E_\btat[B_1(\btat^\dag)] - B_1(\btat) \right] =& \ E[ B_1(\btat) + B_{11}(\btat) ] - \{ B_1(\bta_*) + B_{11}(\bta_*) \} + O(n^{-2}) \non\\
=& \ B_{11}(\bta_*) + O(n^{-2}). \label{eqn:bias2}
\end{align}
Thus, it follows from (\ref{eqn:bias1}) and (\ref{eqn:bias2}) that
$$
E[\widehat{R_3}] = E\left[ 2R_3(\btat) - E_\btat[R_3(\btat^\dag)] + E_\btat[B_1(\btat^\dag)] - B_1(\btat) \right] = R_3(\bta_*) + O(n^{-2}).
$$
Similarly, we show that $E[\widehat{R_4}] = R_4(\bta_*) + O(n^{-2})$.
\hfill$\Box$

\end{document}